# Experimental limits for low-frequency space-time fluctuations from ultrastable optical resonators

S. Schiller[*], C. Lämmerzahl[*,§], H. Müller[+], C. Braxmaier[#], S. Herrmann[+], A. Peters[+]

*Institut für Experimentalphysik, Heinrich-Heine Universität Düsseldorf, Universitätsstr.1, 40225 Düsseldorf, Germany*

+ *Institut für Physik, Humboldt-Universität zu Berlin, Hausvogteiplatz 5-7, 10117 Berlin, Germany*

# *Astrium GmbH, 88039 Friedrichshafen, Germany*

§*Present address: ZARM, Universität Bremen, Am Fallturm, 28359 Bremen, Germany*

**It has been suggested that space-time might undergo fluctuations because of its intrinsic quantum nature. These fluctuations would pose a fundamental limit to the ability of measuring distances with arbitrary precision, beyond any limitations due to standard quantum mechanics. Laser interferometers have recently been proposed as being suited for a search for the existence of space-time fluctuations. Here we present results of a search for space-time fluctuations of very low fluctuation frequencies, in the range from 1 mHz to 0.5 Hz. Rigid optical interferometers made out of sapphire and operated at cryogenic temperature were used. We find an upper limit of $1 \cdot 10^{-24}$ Hz$^{-1}$ for the normalized distance noise spectral density at 6 mHz, and of $1 \cdot 10^{-28}$ Hz$^{-1}$ above 5 mHz, and establish an experimental limit for the parameter of a recently proposed random-walk hypothesis.**

The unification of the theories of gravity and quantum mechanics is one of the outstanding problems of physics. Within a wide range of models and theories attempting to combine the two it is found that space and time exhibit intrinsic spatio-temporal fluctuations. The possibility of existence of observable consequences of these fluctuations, such as decoherence of quantum states or limits to distance measurement precision have been discussed and their detection proposed, e.g by means of particle and light interferometers [1] [2] [3] [4] [5].

Motivated by the concepts of space-time "foam", or "fuzziness", Amelino-Camelia[6] has recently discussed model-independent characteristics of the influence of space-time fluctuations on length measurements. He proposed general forms of the power spectral density $S$ for the relative measurement uncertainty $\Delta L/L$ in a measurement of a length $L$. A hypothesis considering a frequency-independent (white) noise spectrum sets a level $S_{sf}(f) \approx T_p = 5 \cdot 10^{-44}\ Hz^{-1}$, where $T_P$ is the Planck time. Hypotheses with fluctuation strength monotonically increasing with falling frequency were also proposed. One example are random-walk-type fluctuations. Their spectral densities may take different parametrized forms, for example

$$S_{sf}(f) = c^2\ T_p\ \Lambda^{-2}\ f^{-2} = 5 \cdot 10^{-27}\ m^2\ Hz/\Lambda^2\ f^2 \quad (RW1),$$
$$S_{sf}(f) = c^3\ T_p^{2}\ \Lambda^{-3}\ f^{-2} = 7 \cdot 10^{-62}\ m^3\ Hz/\Lambda^3\ f^2 \quad (RW2),$$

where $\Lambda$ is a length scale characteristic of the experiment. The above spectral densities are particular examples of a class of spectral densities that depend on a single phenomenological length parameter and that on dimensional grounds may be written as

$$S_{sf}(f) \approx \left(\frac{\Lambda}{c}\right)\left(\frac{c}{\Lambda}\ T_p\right)^{a}\left(\frac{f}{c/\Lambda}\right)^{g},$$

where $a$, $g$ are arbitrary exponents.[7]

However, since there is no final quantum gravity theory yet, it appears reasonable to view the issue of magnitude and frequency-dependence of the space-time fluctuations as completely open, where theoretical discussions such as the above can be taken as a motivation and guide. It is important to initiate a program aimed at constraining the magnitude and spectral dependence of the hypothetical space-time fluctuations at all experimentally accessible frequencies. In this work we consider, for the first time, the sub-Hz range.

In order to experimentally investigate the hypothesis of space-time fluctuations, Amelino-Camelia[8] has proposed to use the data provided by large-scale Michelson-type laser interferometers developed for gravitational wave detection. In such devices, the fluctuations $\delta(L_1 - L_2)$ of the distance difference of two mirrors from a common beamsplitter is measured. The devices are designed to be sensitive to variations in the metric that occur on timescales that are shorter than the timescale over which the mirrors can be considered to be inertial (freely moving). To implement inertial motion, the mirrors and the beamsplitter are independently supported by pendula whose oscillation period sets the relevant timescale.

The fundamental idea for using gravitational-wave interferometers for searching for space-time fluctuations is that the read-out of such interferometers may exhibit a noise that cannot be explained by standard physics. Furthermore, an important feature of interferometers is that the measurement uncertainty $\Delta L_{min}$ expected from standard physics is not fundamentally dependent on he distances $L_{1,2}$. It is determined by parameters such as seismic, thermal, electronic, and electromagnetic quantum noise levels. Therefore, the achievable measurement uncertainty for detection of length changes *normalized to the arm length*, $\Delta L/L$, can be high in large-scale interferometers, where arm lengths $L_{1,2}$ range from 300 m to 3 km in terrestrial systems and $10^6$ km in space systems.[9]

A fundamental sensitivity limit for detecting length fluctuations arises from quantum fluctuations of the electromagnetic field employed to probe the interferometer. It is described by a white spectral noise $S_{qf} \approx (hc\lambda/4L^2F^2P)$, where $\lambda$ is the wavelength, $F$ is the finesse of each arm's resonator, and $P$ is the power, and $c$ is the speed of light.[10] Power spectral densities close to this limit have been reached experimentally, e.g. $S \sim 2 \cdot 10^{-41}$ Hz$^{-1}$ at fluctuation frequencies $f \sim 10^3$ Hz in the TAMA 300 detector.[11] This level rules out the above hypothesis RW1 for $\Lambda > 15$ km, i.e. exceeding the dimension of the apparatus.

Because mirrors and beamsplitters are independently mounted, on long time scales the lengths $L_{1,2}$ are subject to large drifts due to seismic and thermal disturbances. As a consequence, the large-scale interferometers can only provide limits to length fluctuations at relatively high frequencies $f$ above 10-100 Hz.

In this work we propose to explore the regime of much lower fluctuation frequencies with laser interferometers whose mirrors are rigidly connected. The rigid interferometers envisioned here are configured as cavities (resonators) for optical waves or microwaves. Such rigid interferometers are different from "free-mirror" gravitational interferometers (suspended mass terrestrial gravitational wave interferometers or space interferometers), whose mirrors are not connected. A cavity may be characterized by the resonance frequency of a particular individual longitudinal mode $m$, $\nu^{(m)} = m\, c/2nL$. Here $n$, the index of refraction of the free space between the end mirrors, is equal to unity for a "classical" vacuum but may be regarded as a fluctuating quantity in order to phenomenologically describe fluctuations of space. The distance $L$ between the mirrors is a linear combination of the various inter-atomic distances (bond lengths) $a_i$ of the material to which the mirrors are attached. For a rigid interferometers to be a suitable tool for the detection of quantum space fluctuations, we must assume that its *effective* length $nL$ fluctuates, i.e. that the fluctuations of empty space (fluctuations of $n$) are not compensated for by possible fluctuations of the bond lengths $a_i$. Such a compensation appears unlikely, however, since the physics of light propagation involves the properties of space and of the electromagnetic interaction, but the physics of bond lengths is also determined by the properties of leptonic matter. Experiments with different spacer materials may possibly be helpful in setting limits to compensation effects.

Even for a rigid cavity environmental influences must be kept under control as strongly as possible. Operation of the cavity at cryogenic temperature is favourable since it minimizes changes in cavity length due to unsuppressed temperature changes as well as due to creep (dimensional relaxation). Cryogenic cavities can be operated in a reliable manner over many months [12][13][14] and are therefore in principle suited to probe fluctuation frequencies as low as micro-Hertz. They have also recently been employed to test Lorentz Invariance. [13][14][15][16][17] Notwithstanding the accessibility of the low-frequency window, cryogenic cavities are faced with various technical noise sources, including variations of cavity temperature and tilt, of temperature of the electronic components, and of power and propagation direction of the interrogating electromagnetic wave. Due to the complexity of these effects, in practice it will not be possible to deduce information about space fluctuations of strength below the measured technical noise level. The latter thus yields an upper limit for the strength of space fluctuations. The electromagnetic quantum noise limit described above is a fundamental limit that with enough effort may eventually be reached in some frequency range. Although this limit may in principle be overcome by using non-classical light, the feasibility of testing the hypothesis RW2 for parameter values $\Lambda$ in the meso- or macroscopic range ($\Lambda > 100$ μm) is unlikely, since the allowable powers for stable cryogenic operation must be kept well below 1 mW.

A space fluctuation detector employing cavities can be implemented by an arrangement of two independent cavities. To each cavity a laser is resonantly coupled. In absence of systematic disturbances, under this lock condition the laser frequency is given by $\nu_i = m\, c/2 n_i L_i$, where $i$ is the cavity number. An interference signal between the two waves provides information about the variations $\delta(n_1 L_1 - n_2 L_2)$ of the difference of the measured effective cavity lengths, including any hypothetical quantum space noise. It is a fundamental assumption (also made for Michelson interferometers) that the space fluctuations in the two cavities/arms are uncorrelated, so that the power spectrum of the difference variations $\delta(n_1 L_1 - n_2 L_2)$ can be interpreted as a measure of the power spectrum of the length measurement noise $\Delta(n_i L_i)$ of an individual cavity/arm.

We note here that this assumption would not be required in an alternative scheme in which a single cavity is used whose resonance frequency is compared to an atomic (or molecular) reference frequency. Such comparisons have already been done for tests of fundamental principles [13] [15] [16] but were not analysed in the context of space fluctuations.

Our interferometers are rigid Fabry-Perot optical cavities, each made of a cylindrical sapphire spacer ($L$ = 3 cm long) and two concave mirrors attached to it by molecular adhesion. The cavities are operated at approx. 4 K inside cryostats in order to minimize their thermal expansion coefficient and to allow accurate temperature stabilization. Each cavity is interrogated by a diode-pumped Nd:YAG laser (1064 nm) by coupling the strongly attenuated laser beam spatially into the $TEM_{00}$ cavity mode and detecting the light reflected from the cavity by means of a photodetector. Its electrical output carries information about the detuning between laser frequency and the resonance frequency of the cavity mode. This signal is processed electronically and fed back to an actuator inside the laser that regulates the laser frequency to constantly stay in resonance with the cavity frequency.

The difference $\nu_1(t) - \nu_2(t)$ of the two laser frequencies is obtained by a heterodyne beat between waves split off from each laser and then superposed on a photodetector. The beat frequency is counted using a frequency counter. Its measurement (gate) time implies a corresponding time average of the beat frequency. The measured beat frequency can be approximated as

$$(\nu_1(t) - \nu_2(t))/\nu \approx (dL_{2,sf}/L_2 - dL_{1,sf}/L_1) + (dL_{2,p}/L_2 - dL_{1,p}/L_1) +$$
$$(dL_{2,lock}/L_2 - dL_{1,lock}/L_1) + const,$$

where $dL_{i,sf}(t)$ are the effective length deviations due to quantum space fluctuations, $dL_{i,p}(t)$ are the physical deviations of the cavity length from reference values, $dL_{i,lock}(t)$ denote errors due to drift/fluctuation of the laser frequency lock systems (including the effects of $S_{qf}$), expressed in terms of length. All deviations represent averages over the gate time. $\nu \sim \nu_1 \sim \nu_2 \sim$ 282 THz is the average frequency.

Under the assumption of uncorrelated length fluctuations, one-half the spectral power of the normalized beat frequency $(\nu_1(t) - \nu_2(t))/\nu$, denoted by $S_b$, represents an upper limit for the spectral power $S_{sf}$ due to quantum space fluctuations. Any standard physics noises (described by $dL_{i,p}$ and $dL_{i,lock}$) contribute to the measured spectral density $S_b$ and thus lead to larger estimates for $S_{sf}$.

**Results**

We describe results obtained with two different set-ups. In one set-up (A) [18][19], the two cavities were housed in separate cryostats positioned at a distance of about 2.5 m, with the cavity axes parallel. In the second set-up (B) [14], the two cavities were located orthogonally in a single cryostat, their centers separated by about 10 cm. A good thermal contact between the two cavities was implemented that reduced the sensitivity of the cavity frequency difference to temperature by a factor of about 14 compared to that of an individual cavity frequency.

Fig. 1 shows time records of the beat frequency between the two lasers stabilized to their respective cavities. The data shown for set-up A was taken in October 1999 and is shown after removal of a small linear drift of 1.3 Hz/day, and a 19 Hz frequency jump that occurred at t = 46.6 h due to an accidental bump to one of the cryostats. For set-up B we consider a stable 3-week long section (starting on Jan. 18, 2002) and a relatively "quiet" 14-h section (starting on July 9, 2001) of a year-long run. In the shown B-data, drifts of 53 Hz/day and -46 Hz/day, respectively, were removed.

It can be seen that the data exhibits obvious systematic variations. One reason are drifts of the electronic lock systems. Another reason are periodic disturbances caused by changes in the position of the cavities with respect to the laser beams in consequence of the change in liquid Helium and Nitrogen levels due to evaporation and refill. Nitrogen refills are performed regularly every 3 h, Helium refills approximately every 2 days. In the set-up A such effects are much weaker in part because one of the laser beams was brought to its cavity by means of an optical fiber whose end was fixed relative to the cavity, thereby minimizing relative motion between laser beam and cavity even under varying cryogenic liquid filling.

Frequency measurements were taken every 1 s for setup A and for the 14-h-interval of setup B. The 3-week data of setup B was taken at intervals having a bimodal distribution centered around 1.4 s and 3.6 s, with an average of 2.3 s. The data was therefore averaged over 5 min. The few gaps in the data, including a one-day long gap at $t$=10 d, are removed for simplicity. The data sets obtained in this way represent a time series to which methods of spectral analysis can be applied.[20] Only fluctuation frequencies well below the inverse of the 5 min averaging time are considered for the 3-week B-data.

The spectral density of the cavity frequency (or length) measurement $S_b$ for the three data sets is shown in Fig.2. The highest plotted frequency is 0.5 Hz, one-half of the sampling frequency. The lowest frequency considered is ~ 1 µHz, available from the longer measurement that was possible for setup B. For setups A and B-short we find white spectral densities for frequencies above a few mHz. For the B-short data, the level is much lower, thanks to an increased laser power (more than 1 µW) incident on the cavity, which led to a much higher signal-to-noise of the lock error signal. For all setups a marked increase occurs for lower frequencies, due to the disturbances described above.

The values of spectral noise of setup B-short at higher frequencies together with the values of spectral noise of setup A at the lower frequencies provide experimental limits to the existence of quantum fluctuations of space, independent of specific theoretical hypotheses. Both noise curves are reported here since they arise from setups with different relative orientation and distance between the cavities, implying differing assumptions of noncorrelation of the space fluctuations in each cavity. In terms of $1/f$ - and $1/f^2$ - noise powers, we can set the limits $S_{sf} < 7 \cdot 10^{-31}/f$ (excluding the region above curve (ii)) and $S_{sf} < 3 \cdot 10^{-34}$ Hz/$f^2$ (excluding the region above curve (i)). The spectral densities we obtain are still far from the standard fundamental limits in interferometry described above. In the framework of the RW2 hypothesis proposed by Amelino-Camelia[6], the data of both setups rules out length scales $\Lambda < 0.6$ nm (region above curve (i)). This lower limit is similar to that which can be deduced from the noise floor of the TAMA300 gravitational-wave interferometer at $f = 10^3$ Hz.

**Conclusion**

We have proposed and applied cryogenic resonators to probe the low-frequency region of quantum space fluctuations. Our data yields upper limits for the strength of such fluctuations at frequencies up to $10^8$ times smaller than those probed by terrestrial laser interferometers under development for the detection of gravitational waves. At frequencies around a few µHz, the relative distance noise of space is less than

$S_{sf}= 1·10^{-24}$/Hz, and less than $1·10^{-28}$/Hz above a few mHz. The limit $\Lambda > 0.6$ nm for the RW2 model of Amelino-Camelia was deduced from the low-frequency data.

Our limits were mostly determined by technical effects that we expect to be reduced by several orders using an improved setup. Furthermore, extension to even lower fluctuation frequencies (0.1 µHz) should be feasible; the corresponding longer measurement times pose no fundamental problem. It therefore appears possible that in the near future laboratory-scale experiments will be able to test the hypothesis RW2 for values of $\Lambda$ on the order of the wavelength of light.

Acknowledgements: We thank G. Ruoso and G. Amelino-Camelia for discussions.

FIGURES

Fig. 1

Time traces of the frequency difference $n_1(t) - n_2(t)$ of the resonance frequencies of two optical cavities. Top: setup (A) of two cavities located parallel in separate cryostats, Middle and bottom: setup (B) with two orthogonal cavities in a single cryostat. Each point is a 5 min average of the beat frequency sampled at a rate of approx. 1 Hz.

Fig. 2

Spectral power density $S_b$ of the fluctuations of the measured cavity length $L$ of a single cavity, normalized to $L$. The spectral filters used in estimating the spectra have a bandwidth of 30 µHz for setup A and of 7 µHz and 190 µHz for the long and the short dataset of setup B, respectively. The 95% confidence interval limits of the statistical errors lie 31% below and 81% above the shown values for each setup. Curves (i) and (ii) are spectral powers densities with $1/f^2$ - and $1/f$ - dependence, respectively, that bound the regions excluded by the experimental data.

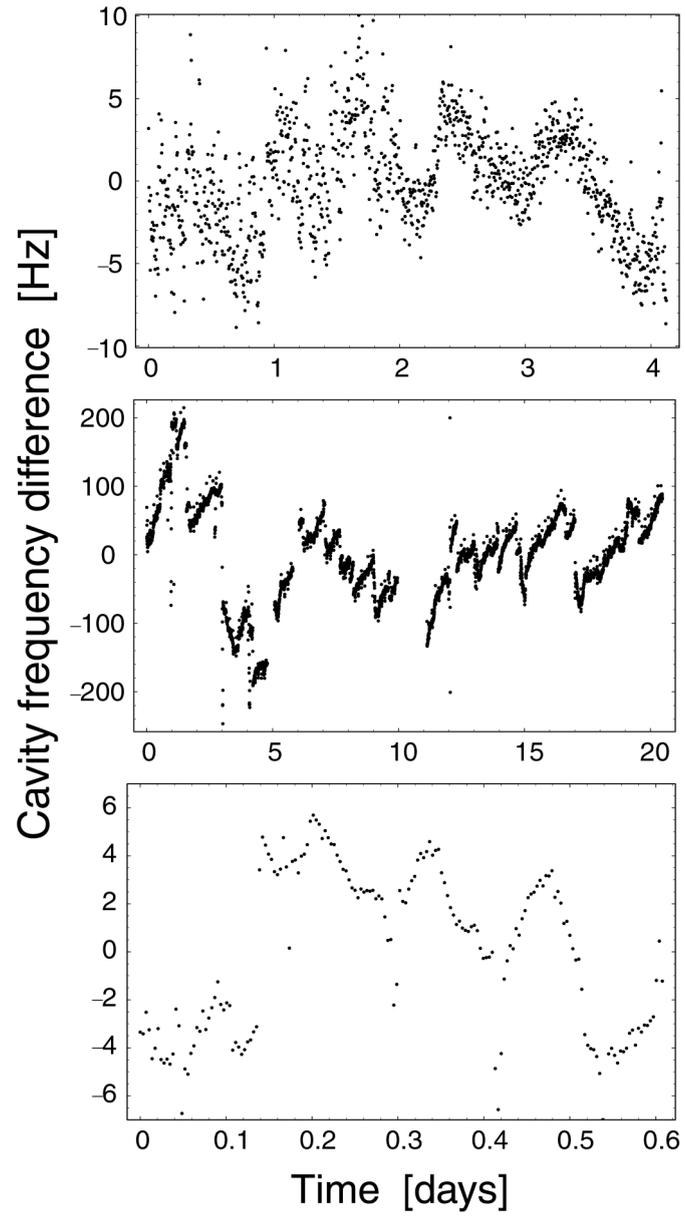

**FIG. 1**

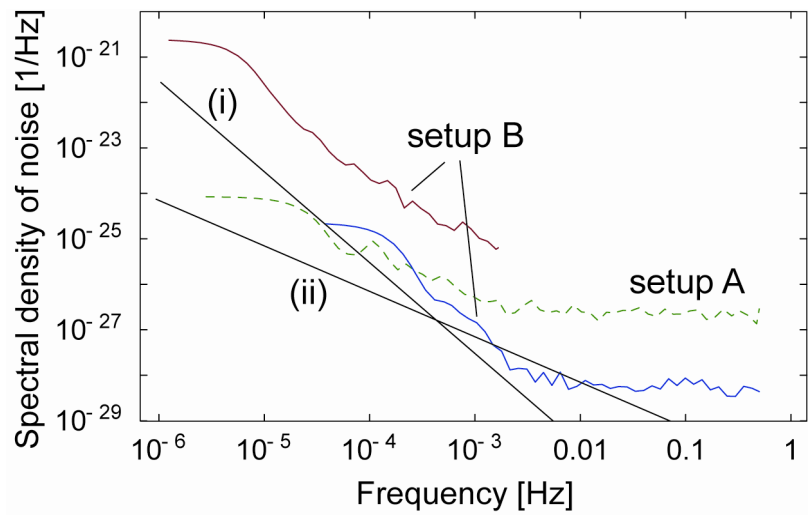

**FIG. 2**